\documentclass[11pt]{article}
\usepackage{moriond,epsfig}
\usepackage{amsmath, amsthm, amsfonts, amssymb}
\usepackage{mathrsfs}
\usepackage{xspace}
\usepackage{cite} 

\newcommand{\beqn}{\begin{eqnarray}}
\newcommand{\eeqn}{\end{eqnarray}}
\newcommand{\beqns}{\begin{eqnarray*}}
\newcommand{\eeqns}{\end{eqnarray*}}

\newcommand{\Kbar    }{\kern 0.2em\overline{\kern -0.2em K}{}\xspace}

\newcommand{\Kz      }{\ensuremath{K^0}\xspace}
\newcommand{\Kzb     }{\ensuremath{\Kbar^0}\xspace}
\newcommand{\KzKzb   }{\ensuremath{\Kz \kern -0.16em \Kzb}\xspace}
\newcommand{\Kp      }{\ensuremath{K^+}\xspace}
\newcommand{\Km      }{\ensuremath{K^-}\xspace}

\newcommand{\KpKm    }{\ensuremath{\Kp \kern -0.16em \Km}\xspace}

\mathchardef\Upsilon="7107
\def\Y#1S{\ensuremath{\Upsilon{(#1S)}}\xspace}

\newcommand{\Tau}{\ensuremath{\tau}\xspace}

\newcommand{\BR}{\ensuremath{{\cal B}}\xspace}

\newcommand{\piz}{\ensuremath{\pi^0}\xspace}

\newcommand{\ee}{\ensuremath{e^+e^-}\xspace}
\newcommand{\mm}{\ensuremath{\mu^+\mu^-}\xspace}
\newcommand{\mmg}{\ensuremath{\mu^+\mu^-(\gamma)}\xspace}

\newcommand{\pp}{\ensuremath{\pi^+\pi^-}\xspace}

\newcommand{\ppg}{\ensuremath{\pi^+\pi^-(\gamma)}\xspace}

\newcommand{\tev}{\ensuremath{\mathrm{\,Te\kern -0.1em V}}\xspace}
\newcommand{\gev}{\ensuremath{\mathrm{\,Ge\kern -0.1em V}}\xspace}
\newcommand{\mev}{\ensuremath{\mathrm{\,Me\kern -0.1em V}}\xspace}
\newcommand{\kev}{\ensuremath{\mathrm{\,ke\kern -0.1em V}}\xspace}
\newcommand{\ev}{\ensuremath{\mathrm{\,e\kern -0.1em V}}\xspace}
\newcommand{\gevc}{\ensuremath{{\mathrm{\,Ge\kern -0.1em V\!/}c}}\xspace}
\newcommand{\mevc}{\ensuremath{{\mathrm{\,Me\kern -0.1em V\!/}c}}\xspace}
\newcommand{\gevcc}{\ensuremath{{\mathrm{\,Ge\kern -0.1em V\!/}c^2}}\xspace}
\newcommand{\mevcc}{\ensuremath{{\mathrm{\,Me\kern -0.1em V\!/}c^2}}\xspace}

\newcommand{\ben}{\begin{enumerate}}
\newcommand{\een}{\end{enumerate}}

\newcommand{\amu}{\ensuremath{a_\mu}\xspace}

\newcommand{\amuhadLO}{\ensuremath{a_\mu^{\rm had,LO}}\xspace}
\newcommand{\amuhadLOpp}{\ensuremath{\amu^{\rm had,LO}[\pi\pi]}\xspace}
\newcommand{\amuhadLOppzz}{\ensuremath{\amu^{\rm had,LO}[\pi\pi2\piz]}\xspace}

\newcommand{\amuSM}{\ensuremath{a_\mu^{\rm SM}}\xspace}

\def\ie{{\it i.e.}}

\newcommand\cf{{\em cf.}\xspace}
\newcommand{\ea}{{\em et al.}\xspace}
\def\@citex[#1]#2{\if@filesw\immediate\write\@auxout{\string\citation{#2}}\fi
  \@tempcnta\z@\@tempcntb\m@ne\def\@citea{}\@cite{\@for\@citeb:=#2\do
    {\@ifundefined
       {b@\@citeb}{\@citeo\@tempcntb\m@ne\@citea
        \def\@citea{,\penalty\@m\ }{\bf ?}\@warning
       {Citation `\@citeb' on page \thepage \space undefined}}%
    {\setbox\z@\hbox{\global\@tempcntc0\csname b@\@citeb\endcsname\relax}%
     \ifnum\@tempcntc=\z@ \@citeo\@tempcntb\m@ne
       \@citea\def\@citea{,\penalty\@m}
       \hbox{\csname b@\@citeb\endcsname}%
     \else
      \advance\@tempcntb\@ne
      \ifnum\@tempcntb=\@tempcntc
      \else\advance\@tempcntb\m@ne\@citeo
      \@tempcnta\@tempcntc\@tempcntb\@tempcntc\fi\fi}}\@citeo}{#1}}

\def\@citeo{\ifnum\@tempcnta>\@tempcntb\else\@citea
  \def\@citea{,\penalty\@m}%
  \ifnum\@tempcnta=\@tempcntb\the\@tempcnta\else
   {\advance\@tempcnta\@ne\ifnum\@tempcnta=\@tempcntb \else
\def\@citea{--}\fi
    \advance\@tempcnta\m@ne\the\@tempcnta\@citea\the\@tempcntb}\fi\fi}
\newenvironment{myquote}
               {\list{}{\leftmargin0cm\indent}%
                \item\relax}
               {\endlist}
\newcommand\allFontSize{\footnotesize}
\newcommand\detailsSize{\allFontSize}
{\begin{myquote}\detailsSize}{\end{myquote}}

\begin{document}
\begin{flushright}
\normalsize
LAL 10-32 \\
\end{flushright}

\vspace*{4cm}
\title{The Current Status of $g-2$}

\author{ B. MALAESCU }

\address{Laboratoire de l'Acc{\'e}l{\'e}rateur Lin{\'e}aire,
         IN2P3/CNRS, Universit\'e Paris-Sud 11, Orsay}

\maketitle\abstracts{
Recently, important updates were made for the hadronic contribution to the theoretical prediction of $g-2$. 
The isospin-breaking-corrections, needed in the comparison of the two pion spectral functions from $\tau$ 
decays and $e^+e^-$ annihilations, were improved using new experimental and theoretical input. 
The recently published BABAR data were included in the global average of $e^+e^-$ spectral functions. 
These data, as well as the ones from $\tau$ decays, were combined using newly developed software, 
featuring improved data interpolation and averaging, more accurate error propagation and systematic 
validation. 
The discrepancy between the $e^+e^-$ and the $\tau$-based result is reduced from previously 2.4 to 1.5 $\sigma$. 
The full Standard Model prediction of $g-2$, obtained using $e^+e^-$ data, differs from the experimental 
value by 3.2 standard deviations. 
}

\section{Introduction}

The Standard Model~(SM) prediction for the anomalous moment of the muon can be conviniently separated into a sum, 
\begin{equation}
\label{alSM}
 a_\mu^{\rm SM} = a_\mu^{\rm QED} + a_\mu^{\rm had} + a_\mu^{\rm weak} , 
\end{equation}
with the QED, hadronic and weak contributions respectively. 
This prediction is limited in precision by the lowest-order hadronic vacuum polarization contribution, 
which together with the hadronic higher order and light-by-light~(LBL) contributions provide the 
total hadronic contribution. 
Owing to unitarity and to the analyticity of the two point correlator, using the optical theorem, 
the lowest order hadronic vacuum polarisation contribution to the anomalous magnetic moment of the muon~(\amuhadLO) 
can be computed through an energy-squared dispersion integral~(ranging from the $\pi^0\gamma$ threshold to infinity):
\begin{equation}
   \label{eq:disp}
   \amuhadLO = \frac{1}{4\pi^3}\int_{m_{\pi^0}^2}^\infty\!\!\!
               ds \, K(s) \, \sigma_{e^+e^-\to{\rm hadrons}}(s)\,,
\end{equation}
where $K(s)$ is a QED kernel function~\cite{br68}. 
Actually, the integration kernel strongly emphasises the low-energy part of the 
spectrum, about 73\% of the lowest order hadronic contribution 
being provided by the $\pi\pi(\gamma)$ final state~\footnote
{
   Throughout this note, final state photon radiation is implied for hadronic final states.
}. 
More importantly, 62\% of its total error stems from the $\pi\pi(\gamma)$ mode, stressing 
the need for ever more precise experimental data in this channel to confirm or not the 
observed deviation between SM prediction and experiment. 

In this note we concentrate on the computation of \amuhadLO. 
We add the other contributions and compare the total SM prediction with the experimental measurement. 
The results presented here were first published in Ref.~\cite{Davier:2009zi}. 
We also recall some previous results from Ref.~\cite{Davier:2009ag}. 

\section{Hadronic data and contributions to \amuhadLO}

A former lack of precision $e^+e^-$-annihilation data inspired the search for an alternative. 
It was found~\cite{adh} in form of $\tau\to\nu_\tau+\pi^-\pi^0,\,2\pi^-\pi^+\pi^0,\,\pi^-3\pi^0$ spectral 
functions~\cite{aleph_old,aleph_new,cleo,opal}, transferred from the charged to 
the neutral state using isospin symmetry. 
During the last decade, new measurements of the $\pi^+\pi^-$ spectral function in $e^+e^-$ annihilation 
with percent accuracy became available~\cite{cmd2,cmd2new,snd,kloe08}, superseding or complementing older and less 
precise data. 

Indeed, fix-energy measurements from the CMD2~\cite{cmd2new} and SND~\cite{snd} experiments at 
the VEPP-2M collider~(Novosibirsk, Russia), achieved comparable statistical errors,
and energy-dependent systematic uncertainties down to $0.8\%$ and $1.3\%$, respectively. 
These measurements have been complemented by results from KLOE~\cite{kloe08}, where a hard-photon ISR technique 
was applied for the first time to precisely determine the \pp cross section between $0.592$ and $0.975\gev$. 
The analysed data sample provides a $0.2\%$ relative statistical error on the \pp contribution to \amuhadLO. 
KLOE does not normalise the $\pp\gamma$ cross section to $\ee\to\mm\gamma$ so that 
the ISR radiator function must be taken from Monte Carlo simulation
(\cf \cite{phokhara} and references therein). 
The  systematic error  assigned to this correction varies between $0.5\%$ and $0.9\%$ (closer to the $\phi$ peak). 
The total assigned systematic error lies between $0.8\%$ and $1.2\%$.

With the increasing precision, which today is on a level with the $\tau$ data in that channel, 
systematic discrepancies in shape and normalisation of the spectral functions were 
observed between the two systems~\cite{dehz02,dehz03}. 
It was found that, when computing the hadronic VP contribution to the muon magnetic anomaly using the $\tau$ 
instead of the \ee data for the $2\pi$ and $4\pi$ channels, the observed deviation 
with the experimental value~\cite{bennett} would reduce to less than $1\sigma$~\cite{md_tau06}. 
The discrepancy between the \Tau and \ee-based predictions decreased after the inclusion of new $\tau$ data 
from the Belle experiment~\cite{belle}, published \ee data from CMD2~\cite{cmd2new} and 
KLOE~\cite{kloe08}~(superseding earlier data~\cite{kloe04}), and a reevaluation of 
isospin-breaking corrections affecting the $\tau$-based evaluation~\cite{Davier:2009ag}.
\footnote
{
   The total size of the isospin-breaking correction to \amuhadLO has been estimated 
   to $(-16.1 \pm 1.9)\cdot 10^{-10}$, which is dominated by the short-distance 
   contribution of $(-12.2 \pm 0.2)\cdot 10^{-10}$~\cite{Davier:2009ag}.
} 
In terms of \amuhadLO, the difference between the $\tau$ and \ee-based evaluations 
in the dominant \pp channel was found to be $11.7 \pm 3.5_{ee} \pm 3.5_{\tau+{\rm IB}}$~\cite{Davier:2009ag} 
(if not otherwise stated, the \amu values are given in units of $10^{-10}$), 
where KLOE exhibits the strongest discrepancy with the $\tau$ data~(without the KLOE 
data the discrepancy reduces from $2.4\sigma$ to $1.9\sigma$). 

Not long time ago the BABAR Collaboration reported~\cite{:2009fg} measurements 
of the processes $\ee\to\ppg, \mmg$ using the ISR method at 10.6\gev centre-of-mass energy.
The detection of the hard ISR photon allows BABAR to cover a large energy range from 
threshold up to $3\gev$ for the two processes. 
The $\pp(\gamma)$ cross section is obtained from the $\pp\gamma(\gamma)$ to $\mm\gamma(\gamma)$ ratio, 
so that the ISR radiation function cancels, as well as additional ISR radiative effects. 
Since FSR photons are also detected, there is no additional uncertainty from radiative corrections at NLO level. 
Experimental systematic uncertainties are kept to 0.5\% in the $\rho$ peak region (0.6--0.9\gev), 
increasing to 1\% outside. 

\section{Combining cross section data}
\label{sec:hvptools}

The requirements for averaging and integrating cross section data are: 
($i$) properly propagate all the uncertainties in the data to the final integral error, 
($ii$) minimise biases, \ie, reproduce the true integral as closely as 
possible in average and measure the remaining systematic error, and 
($iii$) optimise the integral error after averaging while respecting the two 
previous requirements. 
The first item practically requires the use of pseudo-Monte Carlo~(MC) simulation, 
which needs to be a faithful representation of the measurement ensemble~(with all known correlations)
and to contain the full data treatment chain~(interpolation, averaging, integration). 
The second item requires a flexible data interpolation method~(the trapezoidal 
rule is not sufficient as shown below) and a realistic truth model used to test the 
accuracy of the integral computation with pseudo-MC experiments. 
Finally, the third item requires optimal data averaging taking into account all known correlations
to minimise the spread in the integral measured from the pseudo-MC sample.
The details of our newly developed combination procedure~(HVPTools) are given in Ref.~\cite{Davier:2009zi}. 

HVPTools transforms the bare cross section data and associated 
statistical and systematic covariance matrices into fine-grained energy bins, taking 
into account to our best knowledge the correlations within each experiment 
as well as between the experiments~(such as uncertainties in radiative corrections). 
The covariance matrices are obtained by assuming common systematic error sources to be fully correlated. 
To these matrices are added statistical covariances, present for 
example in binned measurements as provided by KLOE, BABAR or the \Tau data, which 
are subject to bin-to-bin migration that has been unfolded by the experiments, thus introducing correlations. 

The interpolation between adjacent measurements of a given experiment uses second order polynomials. 
This is an improvement with respect to the previously applied trapezoidal rule, corresponding to a linear interpolation, which leads
to systematic biases in the integral. 
In the case of binned data, the interpolation function within a bin is renormalised to keep the integral in that
bin invariant after the interpolation. 
The final interpolation function per experiment within its applicable energy domain is discretised into 
small~(1\mev) bins for the purpose of averaging and numerical integration. 

The average weights of the interpolated measurements from different experiments contributing to a given energy bin 
takes into account the precision and the correlations of different measurements and experiments, as well as 
different measurement densities or bin widths within a given energy interval. 
This provides an optimal final uncertainty on \amuhadLO. 

\begin{figure}[bp] \centering 
\includegraphics[width=7.cm]{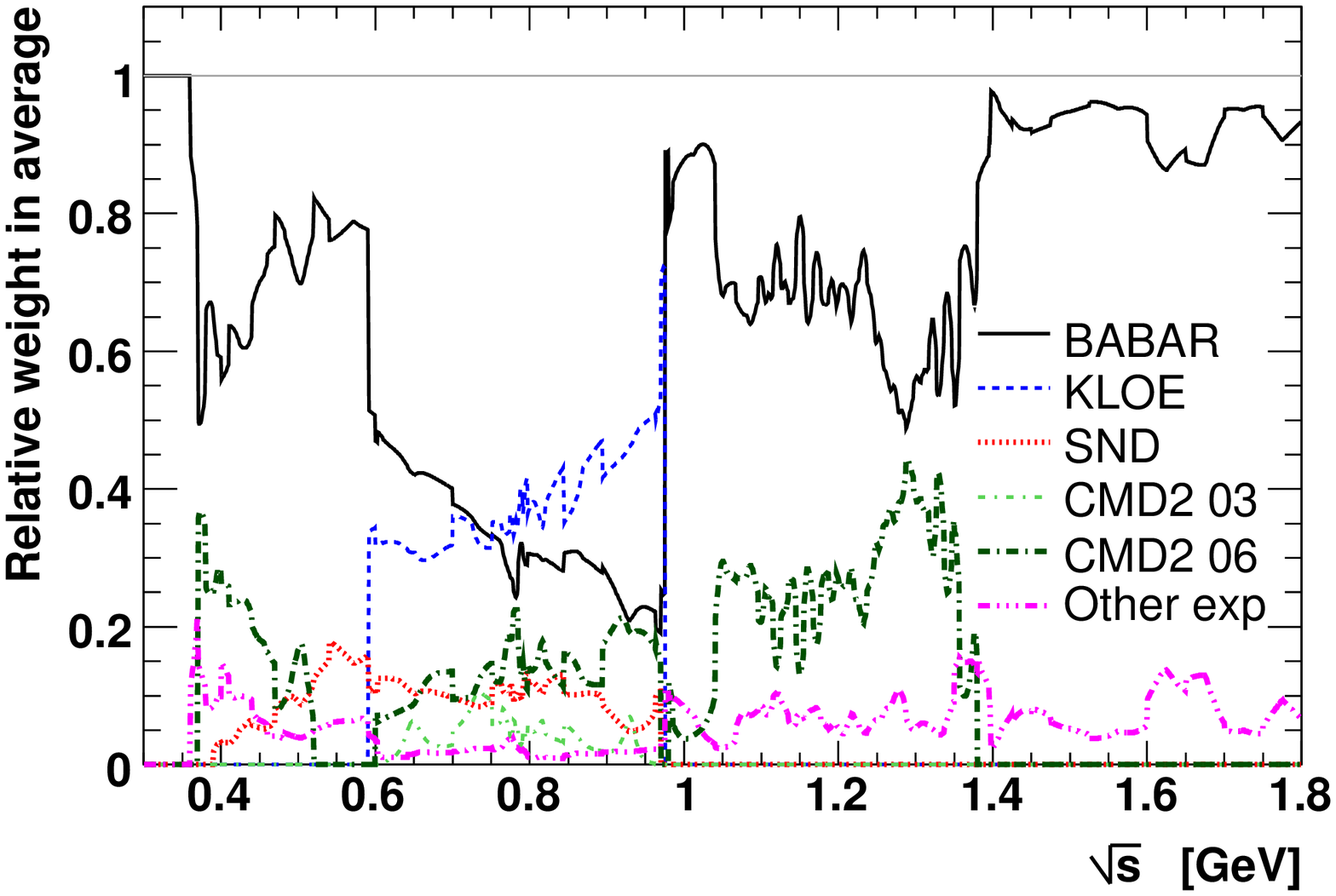}\hspace{1.cm}
\vspace{-0.3cm}
\includegraphics[width=7.cm]{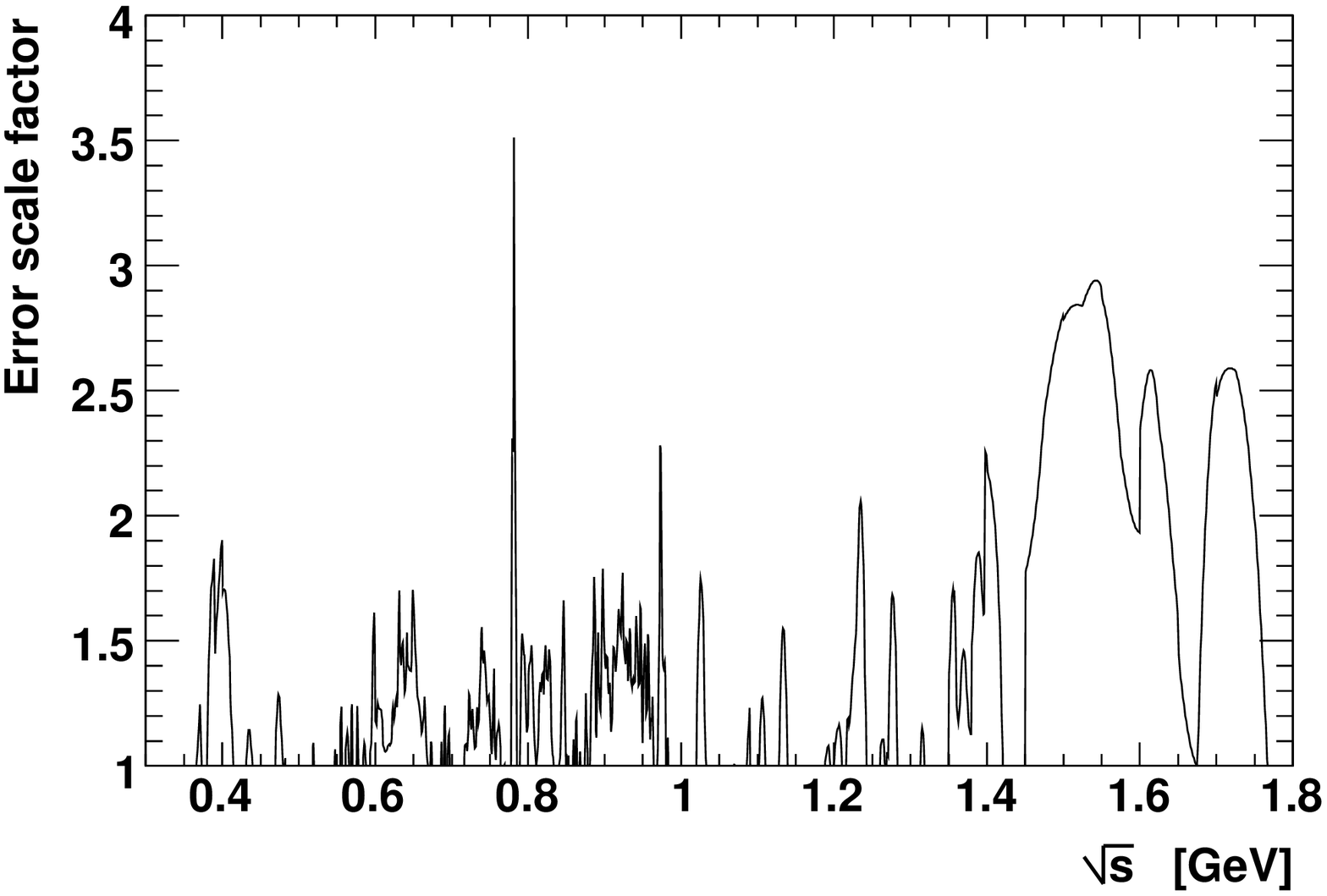}
\caption[.]{\underline{Left:} relative averaging weights per experiment versus $\sqrt{s}$. The experiments labelled ``other exp'' 
                              in the figure correspond to older data with incomplete radiative corrections. 
            \underline{Right:} rescaling factor accounting for inconsistencies among 
             experiments versus $\sqrt{s}$ (see text).} 
\label{fig:weightsAndChi2}
\end{figure}

The left hand plot of Fig.~\ref{fig:weightsAndChi2} shows the weights versus $\sqrt{s}$ the different experiments carry in the average. 
BABAR and KLOE dominate over the entire energy range, completely covered by BABAR. 

If the $\chi^2$ value of a bin-wise average exceeds the number of degrees of freedom ($n_{\rm dof}$), 
the error in this averaged bin is rescaled by $\sqrt{\chi^2/n_{\rm dof}}$ to account for inconsistencies (\cf 
Fig.~\ref{fig:weightsAndChi2}, Right). 
Such inconsistencies frequently occur because most experiments are dominated by systematic uncertainties, which are difficult to estimate. 

We have tested the fidelity of the full analysis chain~(polynomial interpolation, 
averaging, integration) by using as truth representation a Gounaris-Sakurai 
vector-meson resonance model faithfully describing the \pp data. 
The difference between true and estimated \amuhadLO values is a measure for the systematic uncertainty due 
to the data treatment. 
We find negligible bias below $0.1$~(remember the $10^{-10}$ unit), increasing to 0.5~(1.2 without the high-density BABAR data) when 
using the trapezoidal rule for interpolation instead of second order polynomials.

\begin{figure*}[t] \centering 
\begin{center}
\includegraphics[width=7.cm]{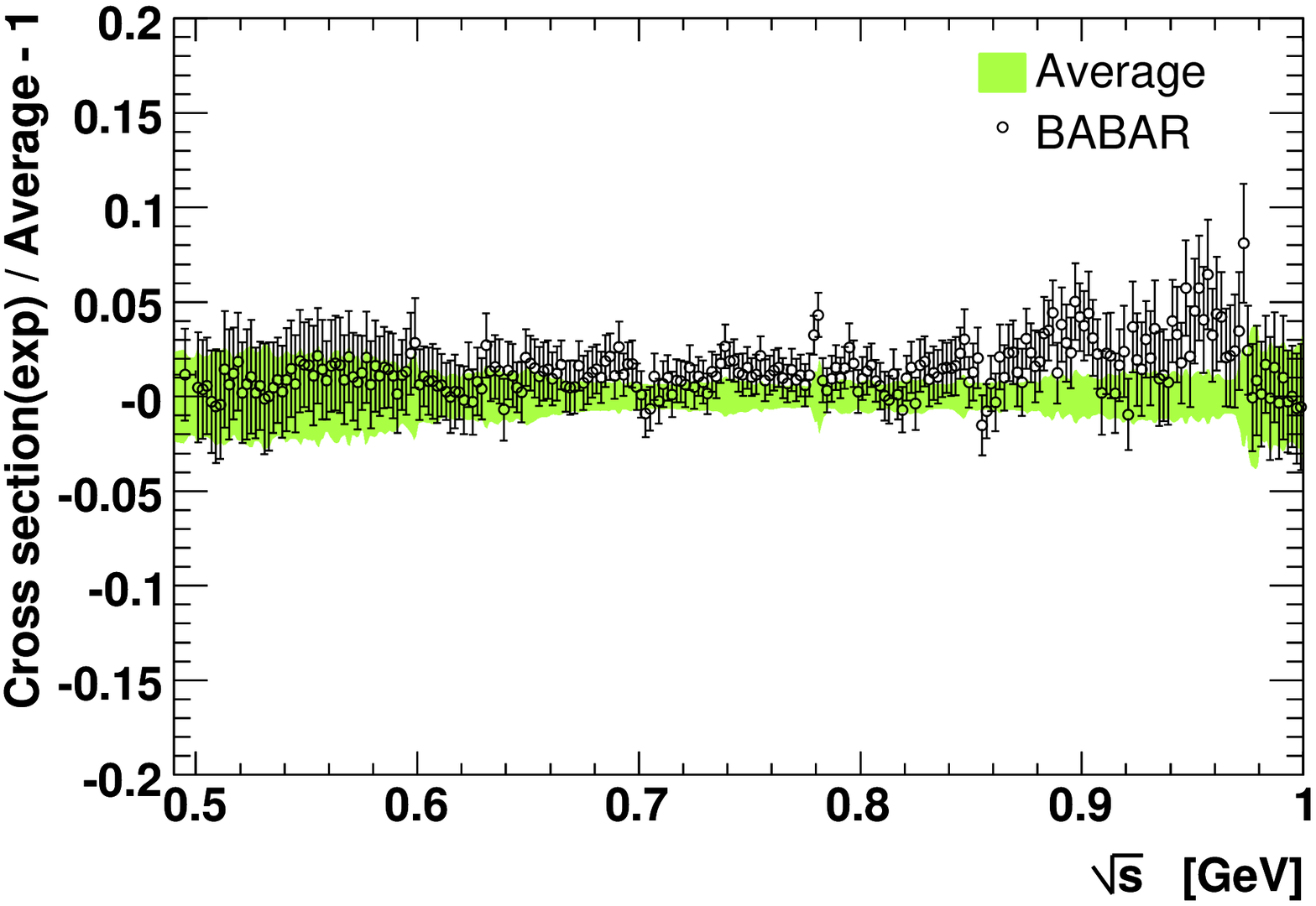}\hspace{1.cm}
\includegraphics[width=7.cm]{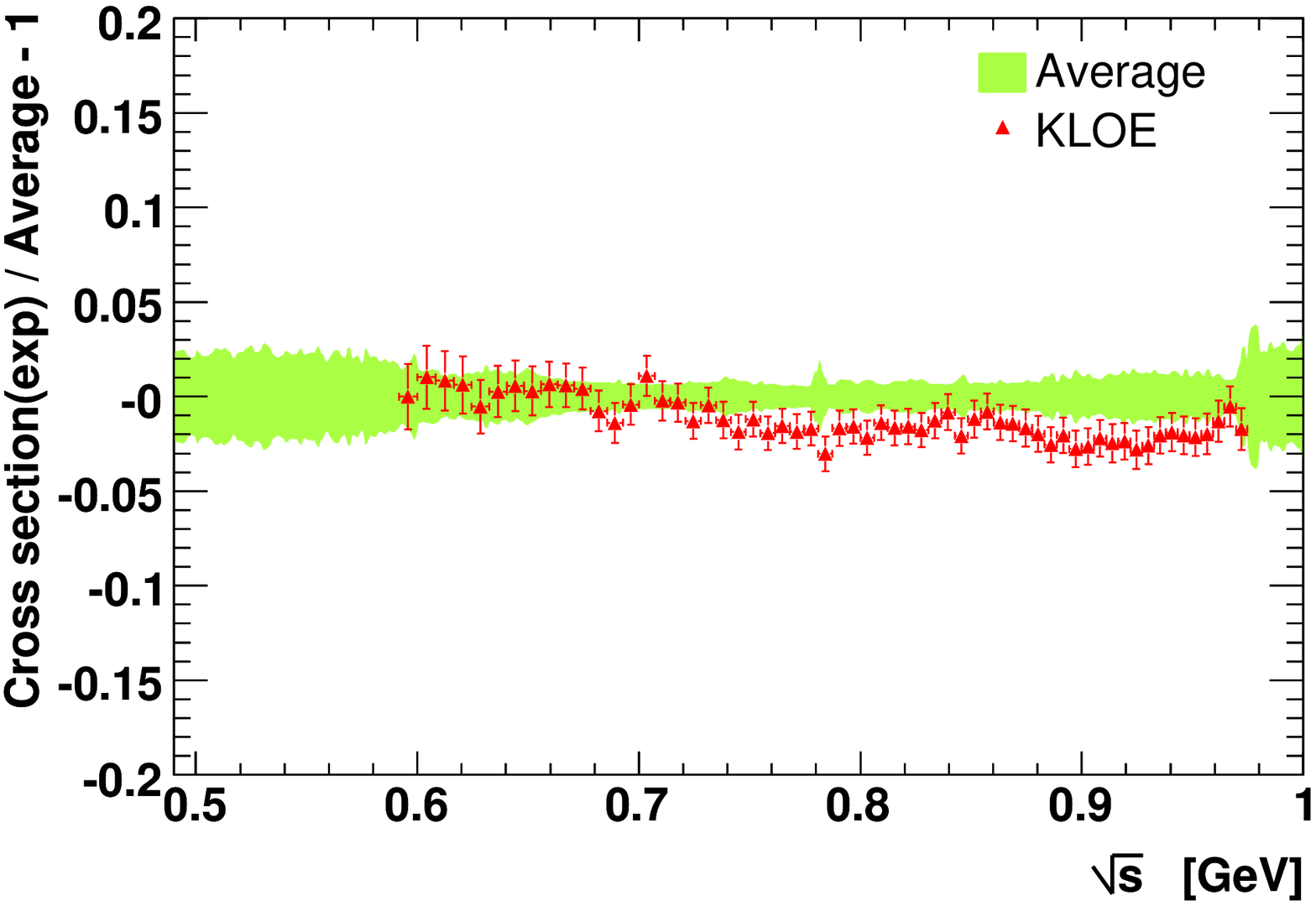}
\end{center}
\vspace{-0.3cm}
\caption[.]{Relative cross section comparison between individual experiments 
           (symbols) and the HVPTools average (shaded band) computed from all 
           measurements considered. Shown are BABAR~(left) and KLOE~(right). 
}
\label{fig:comp}
\end{figure*}

Fig.~\ref{fig:comp} shows the relative 
differences between BABAR, KLOE, and the average. 
Fair agreement is observed, though with a tendency to larger (smaller) cross sections above $\sim$$0.8\gev$ for BABAR~(KLOE). 
It is interesting to point out that the same type of tendency shows up when comparing the KLOE measurement with 
the IB-corrected $\tau$ average~\cite{Davier:2009ag}. 
These inconsistencies (among others) lead to the error rescaling shown 
versus $\sqrt{s}$ in Fig.~\ref{fig:weightsAndChi2}, Right. 

\begin{table}[t]
\vspace{-0.15cm}
  \caption{\label{tab:results}
    Evaluated $\amuhadLOpp$ contributions from the \ee data 
    for different energy intervals and experiments. Where two errors are given, the 
    first is statistical and the second systematic. We also recall 
    the \Tau-based result.  
}
\begin{center}
{ \scriptsize 
\begin{tabular}{ccc} 
\hline\noalign{\smallskip}
  Energy range (GeV)  &  Experiment   &  $\amuhadLO[\pi\pi]$ ($10^{-10}$) \\
\noalign{\smallskip}\hline\noalign{\smallskip}
$2m_{\pi^\pm}-0.3$ & Combined \ee (fit)  & $0.55\pm 0.01$ \\ 
$0.30-0.63$   & Combined \ee  & $132.6 \pm 0.8 \pm 1.0$ ($1.3_{\rm tot}$) \\
$0.63-0.958$ & CMD2 03        & $361.8 \pm 2.4 \pm 2.1$ ($3.2_{\rm tot}$) \\
             & CMD2 06        & $360.2 \pm 1.8 \pm 2.8$ ($3.3_{\rm tot}$) \\
             & SND  06        & $360.7 \pm 1.4 \pm 4.7$ ($4.9_{\rm tot}$) \\
             & KLOE 08        & $356.8 \pm 0.4 \pm 3.1$ ($3.1_{\rm tot}$) \\
             & BABAR 09       & $365.2 \pm 1.9 \pm 1.9$ ($2.7_{\rm tot}$) \\
             & Combined \ee   & $360.8 \pm 0.9 \pm 1.8$ ($2.0_{\rm tot}$) \\
$0.958-1.8$  & Combined \ee   &  $14.4 \pm 0.1 \pm 0.1$ ($0.2_{\rm tot}$) \\ 
\noalign{\smallskip}\hline\noalign{\smallskip}
Total        & Combined \ee   & $508.4 \pm 1.3 \pm 2.6$ $(2.9_{\rm tot})$\\
Total        & Combined \Tau  & $515.2\pm 2.0_{\rm exp}\pm 2.2_\BR\pm 1.9_{\rm IB}$ $(3.5_{\rm tot})$ \\
\noalign{\smallskip}\hline
\end{tabular}
}
\end{center}
\end{table}

A compilation of results for \amuhadLOpp for the various sets of experiments and energy 
regions is given in Table~\ref{tab:results}. 
The comparison with our previous result~\cite{Davier:2009ag}, $\amuhadLOpp=503.5\pm3.5_{\rm tot}$, 
shows that the inclusion of the new BABAR data significantly increases the central value 
of the integral, without however providing a large error reduction. 
This is due to the incompatibility between mainly BABAR and KLOE, causing an increase of
the combined error. 
In the energy interval between 0.63 and 0.958\gev, the discrepancy 
between the \amuhadLOpp evaluations from KLOE and BABAR amounts to $2.0\sigma$.
Using only the BABAR data to evaluate \amuhadLOpp one finds~\cite{:2009fg} $514.1 \pm 2.2_{\rm stat} \pm 3.1_{\rm syst}$, 
which is in very good agreement with the result from the \Tau average.
Including BABAR in the global average reduced the difference between the $\tau$ and $\ee$-based predictions to $1.5\sigma$. 
A difference~(slope) in the shape of the two spectral functions subsists however\cite{Davier:2009zi}, mainly due to the difference 
between the \Tau average and KLOE measurement. 

We also reevaluate the $\ee\to\pp2\piz$ contribution to \amuhadLO. 
The CMD2 data used previously~\cite{cmd2pp2pi0} have been superseded by modified or 
more recent, but yet unpublished data~\cite{logashenko}, recovering agreement with the 
published SND cross sections~\cite{sndpp2pi0}. Since the new data are unavailable, we 
discard the obsolete CMD2 data from the $\pp2\piz$ average, finding 
$\amuhadLOppzz=17.6\pm0.4_{\rm stat}\pm1.7_{\rm syst}$ (compared to 
$17.0 \pm 0.4_{\rm stat} \pm 1.6_{\rm syst}$ when including the obsolete CMD2 data). 

Adding to the \ee-based \amuhadLOpp and \amuhadLOppzz results the remaining exclusive 
multi-hadron channels as well as perturbative QCD~\cite{md_tau06}, we find for the 
complete lowest order hadronic term $\amuhadLO[\ee] = 695.5 \pm 4.0_{\rm exp}\pm  0.7_{\rm QCD}~(4.1_{\rm tot})\,.$
It is noticeable that the error from the \pp channel now equals the one from 
all other contributions to \amuhadLO.
Adding further the contributions from higher order hadronic loops, hadronic light-by-light scattering, as well as QED and electroweak effects,
we obtain the SM prediction~(still in $10^{-10}$ units)
\beqns
  \amuSM[\ee] &=& 11\,659\,183.4 \pm 4.1 \pm 2.6 \pm 0.2~(4.9_{\rm tot})\,,
\eeqns
where the errors have been split into lowest and higher order hadronic, and 
other contributions, respectively. 
The $\amuSM[\ee]$ value deviates from the experimental average by $25.5 \pm 8.0$ ($3.2\sigma$).

\begin{figure}[htbp] \centering 
\includegraphics[width=6.4cm]{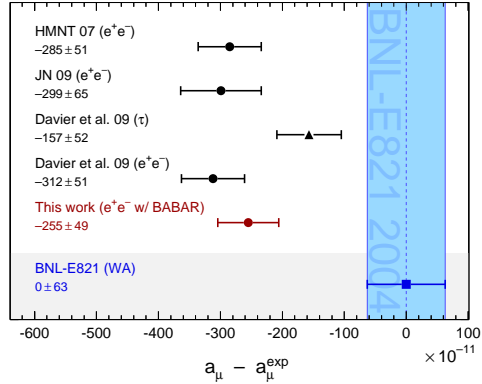}
\vspace{-0.4cm}
\caption{Compilation of recent results for $\amuSM$,
        subtracted by the central value of the experimental average~\cite{bnl}.
        The shaded vertical band indicates the experimental error. 
        The SM predictions are taken from: HMNT 07~\cite{hmnt}, JN 09~\cite{jeger},
        Davier \ea 09~\cite{Davier:2009ag} (\Tau-based and \ee including KLOE), 
        and the \ee-based value from this work. }
\label{fig:amures}
\end{figure}
A compilation of recent SM predictions for \amu compared with the experimental
result is given in Fig.~\ref{fig:amures}. The BABAR results are not yet contained
in evaluations preceding the present one. The result by HMNT~\cite{hmnt} contains 
older KLOE data~\cite{kloe04}, which have been superseded by more recent 
results~\cite{kloe08}, leading to a slightly larger value for \amuhadLO.

\section{Conclusions} 

We have presented a reevaluation of the lowest order hadronic contribution to the muon 
magnetic anomaly in the dominant \pp channel, using new precision data published 
by the BABAR Collaboration. After combination with the other \ee  data a 
$1.5\sigma$ difference with the \Tau data remains for the dominant \pp contribution. 
For the full \ee-based Standard Model prediction, including also a reevaluated
$\pp2\piz$ contribution, we find a deviation of $3.2\sigma$ from experiment
(reduced from $3.7\sigma$ without BABAR). The deviation reduces to $2.9\sigma$ 
when excluding KLOE data, and further decreases to $2.4\sigma$ when using only 
the BABAR data in the \pp channel. As a reminder, the \Tau-based result deviates
by $1.9\sigma$ from the Standard Model.

The present situation for the evaluation of \amuhadLOpp is improved compared to 
that of recent years, as more input data from quite different experimental facilities 
and conditions have become available: \ee energy scan, \ee ISR from low and high energies, 
\Tau decays. Our attitude has been to combine all the data and include in the 
uncertainty the effects from differences in the spectra. At the moment the ideal 
accuracy cannot be reached as a consequence of the existing discrepancies due to 
uncorrected or unaccounted systematic effects in the data. A critical look must
be given to the different analyses in order to identify their weak points and to 
improve on them or to assign larger systematic errors.

Problems also persist in the $\pp2\piz$ mode, where the \Tau
and \ee-based evaluations differ by $(3.8\pm2.2)\cdot10^{-10}$, but also the 
\ee data among themselves exhibit discrepancies. 
Fortunately, new precision data from BABAR should soon help to clarify the situation in that channel. 

\section*{Acknowledgments}

The author would like to thank M.~Davier, A.~H\"ocker, G.~L\'opez Castro, X.H.~Mo, G.~Toledo S\'anchez, P.~Wang, C.Z.~Yuan and Z.~Zhang for their 
fruitful collaboration.

\end{document}